\documentclass[12pt,twoside,a4paper]{article}
\usepackage{amsmath}
\usepackage{epsfig}
\usepackage{subfigure}

\font\bit=cmbxti10 

\begin{document}

\title{ 
\vskip-3cm 
{\normalsize
\hfill MZ-TH/06-27\\
}
\vskip4cm 
{\bf QCD condensates}\\
{\bf of dimension $D=6$ and $D=8$}\\
{\bf from hadronic $\tau$-decays}
\\[3ex] 
}

\date{}
\author{ 
A.~A.~Almasy\footnote{Email: almasy@thep.physik.uni-mainz.de},~ 
K.~Schilcher,~ 
H.~Spiesberger\footnote{Email: hspiesb@thep.physik.uni-mainz.de} \\[1ex]
\normalsize{Institut f\"ur Physik, Johannes-Gutenberg-Universit\"at,}\\ 
\normalsize{Staudinger Weg 7, D-55099 Mainz, Germany} \\[10ex]
}

\maketitle

\begin{abstract}
\medskip 
\noindent 
The high-precision data from hadronic $\tau$ decays allows one to
extract information on QCD condensates. Using the finalized ALEPH data,
we obtain a more rigorous determination of the dimension 6 and 8
condensates for the $V-A$ correlator. In particular, we find that the
recent data fix a certain linear combination of these QCD condensates to
a precision at the level of $O(2)\,\%$. Our approach relies on more
general assumptions than alternative approaches based on finite energy
sum rules.
\end{abstract}

\thispagestyle{empty}

\clearpage  

\section{Introduction}

The physics of hadronic $\tau$ decays has been an important testing
ground for QCD since long. The precise data from the LEP collaborations
have allowed us to obtain invaluable information on fundamental and
effective parameters of the theory. With the completion of the final
analysis by the ALEPH collaboration \cite{aleph-final} it is timely to
exploit these data also for a more precise extraction of QCD condensate
parameters.

In a previous letter \cite{CSSS} we had used earlier ALEPH data
\cite{aleph} in a functional method to extract within rather general
assumptions the QCD condensate of dimension $D=6$ related to the $V-A$
current. Here we report the results of a reanalysis using the more
precise final data from ALEPH. Moreover, we found that the higher
quality of the experimental data allows us also to obtain information on
the condensates of higher dimension. In particular, we find that the
data imply a strong correlation between $D=6$ and $D=8$ $V-A$
condensates.

We extract the condensates from a comparison of the time-like
experimental data with the asymptotic space-like results from theory.
The assumptions of our approach are quite general. The essential
property of the exact correlator in the space-like region on which our
approach relies is that its fall-off with the 4-momentum squared $s$ is
determined by at most two lowest-dimension terms in the operator product
expansion (OPE). When we aim at a determination of the $D=6$ condensate,
we assume a fall-off like $1/s^3$ within an error band that scales like
$1/s^4$. In the analysis where we want to determine the $D=6$ and $D=8$
condensates, we correspondingly assume a fall-off like a sum of $1/s^3$
and $1/s^4$ terms within an error band described by a $1/s^5$ term.
These assumptions are more general than those of other approaches since
they do not refer to the positive axis in the $s$-plane. Furthermore our
assumptions are quite independent of perturbation theory or indeed of
QCD itself. Similarly to the Weinberg sum rules, they only depend on the
pattern of chiral symmetry breaking.  Of course, as QCD is well
established, we like to discuss our results in the language of QCD and
the OPE. For this reason we also include the $O(\alpha_{s})$ correction
to the $1/s^{3}$ term. If it was available, we could have included the
perturbative correction to the $1/s^4$ term as well. However, as we will
see, the inclusion of higher-order terms does not qualitatively affect
our results.

One should, however, keep in mind that it is not possible (not even in
principle) to reconstruct the correlator in the space-like region from
error afflicted time-like data, as this constitutes an analytic
continuation from a finite domain. One has to stabilize this ``ill-posed
problem'' by suitable additional assumptions. For example, the
application of finite energy sum rules (FESR) for an extraction of QCD
condensates could be justified if the result of QCD and the operator
product expansion in the space-like region is simply a series in powers
of $1/s$ times condensates (vacuum expectation values of operators) and
if there are no truly non-perturbative terms. In this case, each moment
would pick out a single operator. However, this is unfortunately not the
case since logarithms arising within perturbation theory do not fall in
this class of functions. The higher and lower dimensional condensates
contribute to a given moment starting with the inclusion of corrections
of order $O(\alpha_{s}^{2})$ \cite{Launer}. It is also known that the
perturbation series starts to diverge at some, not very high, order. As
higher orders of the relevant Wilson coefficients are not known, one can
only hope that the extraction of the condensates is stable when
correction terms are included. There is the additional problem of the
contribution of the truly non-perturbative terms (non-OPE) to the
integral on the circle in the complex plane, even if this uncertainty,
which is expected to be most prominent near the physical cut, can be
reduced by choosing suitable linear combinations of moments.

We should expect that there is a price to be paid for the generality of
assumptions of our approach. However, even if it turns out that the
errors in the values of the extracted parameters are relatively large,
we can hope that our results lend additional confidence to the numerical
results obtained with the help of FESR.


\section{QCD condensates}

We consider the polarization operator of hadronic vector and
axial-vector charged currents, $J_{\mu }=V_{\mu }=\bar{u}\gamma _{\mu
  }d$ and $J_{\mu }=A_{\mu }=\bar{u}\gamma _{\mu }\gamma _{5}d$,
\begin{eqnarray}
\Pi _{\mu \nu }^{J} 
& = & i\int dxe^{iqx}\langle TJ_{\mu }(x)J_{\nu}(0)^{\dagger}\rangle 
\\
& = & \left( -g_{\mu \nu } q^{2} + q_{\mu }q_{\nu } \right) 
\Pi_{J}^{(0+1)}(q^{2}) + g_{\mu \nu } q^2 \Pi _{J}^{(0)}(q^{2}) \,.
\nonumber
\end{eqnarray}
The conservation of the vector current implies $\Pi _{V}^{(0)}=0$. The
connection to experimental observables is most easily expressed with the
help of the spectral functions which are related to the absorptive part
of the correlators. Keeping the normalization as defined in most of the
experimental publications, the functions
\begin{equation}
v_{j}(s) = 2\pi {\rm Im}\Pi _{V}^{(j)}(s),~~~~a_{j}(s)
         = 2\pi {\rm Im}\Pi_{A}^{(j)}(s) 
\end{equation}
can be extracted from the decay spectrum of hadronic $\tau $-decays. We
restrict the present study to the $V-A$ component which is related to
the branching ratios of $\tau$ decays,
\begin{equation}
R_{\tau, {\rm V-A}} = 
\frac{B(\tau \rightarrow \nu _{\tau } + {\rm hadrons},~{\rm V-A})}%
     {B(\tau \rightarrow \nu _{\tau }+e+\bar{\nu}_{\tau })}
\, ,
\label{rtauvma} 
\end{equation}
through
\begin{eqnarray}
v_1(s) - a_1(s) & = & 
\frac{m_{\tau}^2}{6\left\vert V_{ud}\right\vert ^{2}S_{{\rm EW}}}
\frac{{\rm d}R_{\rm V-A}}{{\cal B}_e {\rm d} s}
\left(1-\frac{s}{m_{\tau}^{2}}\right)^{-2} 
\left(1 + 2\frac{s}{m_{\tau }^{2}}\right)^{-1} \, ,
\\
a_0(s) & = & 
\frac{m_{\tau}^2}{6\left\vert V_{ud}\right\vert ^{2}S_{{\rm EW}}}
\frac{{\rm d}R_{\rm A}}{{\cal B}_e {\rm d} s}
\left(1-\frac{s}{m_{\tau}^{2}}\right)^{-2} 
\,. 
\label{specfun-a0}
\end{eqnarray}
Here, ${\cal B}_e = 0.17810 \pm 0.00039$ \cite{aleph-final} is the
branching fraction of the electron channel, $V_{ud}$ is the CKM-matrix
element, $\left| V_{ud} \right| = 0.9746 \pm 0.0006$ \cite{vud}, the
$\tau$ mass is $m_{\tau} = 1.777$ GeV and $S_{\rm EW} = 1.0198 \pm
0.0006$ accounts for electroweak radiative corrections \cite{sew}.  The
spin-0 axial vector contribution $a_{0}(s)$ is dominated by the one-pion
state, $a_{0,\pi}(s)=2\pi ^{2}f_{\pi}^{2}\delta (s-m_{\pi }^{2})$, with
the $\pi$-decay constant $f_{\pi} = 0.1307$ GeV \cite{PDG}.

We use the final experimental data from the ALEPH collaboration
\cite{aleph-final} because they have the smallest experimental
errors\footnote{Earlier measurements have been reported in \cite{opal},
  see also \cite{Roney} and references given there.}.  They are given by
binned and normalized event numbers related to the differential
distribution $dR_{\tau,V,A}/ds$. The quality of these data has increased
a lot as compared to those of Ref.\ \cite{aleph} which we used in our
previous analysis \cite{CSSS}: first, the higher statistics has allowed
the ALEPH collaboration to double the number of bins, and, secondly, the
experimental errors decreased due to both higher statistics and a better
understanding of systematic errors.

The $(V-A)$ correlator is special since it vanishes identically in the
chiral limit ($m_{q}=0)$ to all orders in QCD perturbation theory.
Renormalon ambiguities are thus avoided.  We will neglect in our
analysis perturbative contributions proportional to the light quark
masses.  Non-perturbative terms can be calculated for large $\left\vert
  s\right\vert $ by making use of the operator product expansion of QCD
\begin{equation}
\Pi_{\rm V-A}^{(0+1)}(s) = \sum_{D\geq 6} 
\frac{{\cal O}_{D}^{\rm V-A}}{(-s)^{D/2}}
\left(1 + c_{D}^{\rm NLO}\frac{\alpha_s(\mu^2)}{\pi} 
+ O(\alpha_s^2) \right)
\label{ope}
\end{equation}
where ${\cal O}_{D}^{\rm V-A}$ are vacuum matrix elements of local
operators of dimension $D$. For the correlator with spin $0+1$ and in
the chiral limit, the sum begins with terms of dimension $D=6$. Assuming
vacuum dominance or the factorization approximation which holds, e.g.,
in the large-$N_C$ limit, the matrix elements can be written as
\begin{eqnarray}
{\cal O}_6^{\rm V-A} & = & 
-\frac{64\pi}{9}\alpha_s\langle\bar qq\rangle^2 \, ,
~~~~~ c_6^{\rm NLO} = 
\frac{1}{4}
\left[\tilde{c}_6+\mbox{ln}\left(\frac{\mu^2}{-s}\right)\right] \, ;
\nonumber
\\[3ex]
{\cal O}_8^{\rm V-A} & = & 
-4\pi\alpha_s i \left(1 - N_C^{-2}\right)
 \langle\bar q q\rangle 
 \langle \bar q \gamma^{\alpha \beta} G_{\alpha\beta} q \rangle \, ;
\\[3ex]
{\cal O}_{10}^{\rm V-A} & = & 
- \frac{8}{9}\pi \alpha_s \langle\bar qq\rangle^2 
\left( \frac{50}{9} m_0^4 + 32 \pi \alpha_s 
\langle G_{\alpha\beta}G^{\alpha\beta}\rangle \right) \, .
\nonumber
\label{cond468}
\end{eqnarray}
The complete expression for ${\cal O}_6^{\rm V-A}$ is known to involve
two operators \cite{nlo-coeffs}. However, our analysis does not rely on
the factorization hypothesis since we are going to determine ${\cal
  O}_6^{\rm V-A}$ and ${\cal O}_8^{\rm V-A}$, but not the condensates
$\langle\bar qq\rangle$ or $\langle \bar q \gamma_{\alpha \beta}
G^{\alpha \beta} q \rangle$ separately. In the result for ${\cal
  O}_{10}^{\rm V-A}$ \cite{ZyabD10}, $m_0^2$ is defined through the
5-dimensional quark-gluon mixed condensate.  Starting from the second
order, coefficients in perturbation theory depend on the regularisation
scheme implying that the values of the condensates are scheme-dependent
quantities. The NLO corrections for ${\cal O}_6^{\rm V-A}$ were computed
first in \cite{CSGL} and the coefficient $\tilde{c}_6$ was found equal
to $247/12$. This calculation was based on the BM definition of
$\gamma_5$ in dimensional regularisation. A different treatment of
$\gamma^5$ as used in Ref.\ \cite{AC} leads to $\tilde{c}_6=89/12$.

The typical scales determining the condensates are around 300 MeV. For
example, from the Gell-Mann--Oakes--Renner relation \cite{GMOR} one has
$\langle \bar{q}q\rangle \simeq -(250 \, {\rm MeV})^{3}$, and from
charmonium sum rules one expects ($\alpha _{s}/\pi )\langle G^{2}\rangle
\simeq (300 \, {\rm MeV})^{4}$ \cite{SVZ}. Taking the results from FESR
approaches as a guide, we expect also ${\cal O}_8^{\rm V-A}$ and ${\cal
  O}_{10}^{\rm V-A}$ to be of order $10^{-3}$ GeV$^8$ and $10^{-3}$
GeV$^{10}$, resp.\ (see e.g.\ \cite{Ioffe}).  This is small enough so
that the OPE makes sense.  If ${\cal O}_{8,10}^{\rm V-A}$ would be much
larger, radiative corrections to higher-dimension condensates would mix
significantly with the lower-dimension ones through their imaginary
parts.

In the following we shall summarize the functional method\footnote{The
  functional method underlying our analysis has first been described in
  \cite{auberson1,auberson2}.} introduced in Ref.\ \cite{CSSS} which
allows, in principle, to extract the condensates from a comparison of
the data with the asymptotic space-like QCD results under rather general
assumptions.


We consider a set of functions $F(s)$ (to be identified with $\Pi_{\rm
  V-A}^{(0+1)}(s)$) expressed in terms of some squared energy variable
$s$ which are admissible as a representation of the true amplitude if
\begin{itemize}
\item[\bit i)] $F(s)$ is a real analytic function in the complex
  $s$-plane cut along the time-like interval $\Gamma _{R}=[s_{0},\infty
  )$. The value of the threshold $s_{0}$ depends on the specific
  physical application ($s_{0}=(2m_{\pi })^{2}$ for $\Pi_{V}$,
  $s_{0}=m_{\pi }^{2}$ for $\Pi_{A}$).
\item[\bit ii)] The asymptotic behavior of $F(s)$ is restricted by
  fixing the number of subtractions in the dispersion relation between
  $F(s)$ and its imaginary part along the cut $f(s)={\rm
    Im}F(s+i0)|_{s\in \Gamma _{R}}$ (for $\Pi_{\rm V-A}^{(0+1)}(s)$ no
  subtractions are needed):
  \begin{equation}
  F(s)=\frac{1}{\pi }\int_{0}^{\infty }\frac{f(x)}{x-s}dx
  + \frac{f_{\pi}^2}{s}\,,
  \label{disprel}
  \end{equation}
  where the term with $f_{\pi}^2$ accounts for the contribution from the
  pion pole, $a_0$ (Eq.\ \ref{specfun-a0}), so that $f(x)$ can be taken
  directly from the ALEPH data.
\end{itemize}

We have two sources of information which will be used to determine
$F(s)$ and $f(s)$. First, there are experimental data in a {\sl
  time-like interval} $\Gamma_{\rm exp} = [s_0, s_{\rm max}]$ with $s_0
> 0$ for the imaginary part of the amplitude. Although these data are
given on a sequence of adjacent bins, we describe them by a function
$f_{{\rm exp}}(s)$. We assert that $f_{{\rm exp}} $ is a real, not
necessarily continuous function. The experimental precision of the data
is described by a covariance matrix $V(s, s^{\prime})$.

On the other hand, we have a theoretical model, in fact QCD. From
perturbative QCD we can obtain a prediction for the amplitude in a {\sl
  \ space-like interval} $\Gamma_L = [s_2, s_1] $. This model amplitude
is a continuous function of real type, but does not necessarily conform
to the analyticity property {\bit i)}. Since perturbative QCD is
expected to be reliable for large energies, we expect that there is also
useful information about the imaginary part of the amplitude provided
that $|s|$ is large, i.e.\ we can also use QCD predictions for $f(s) =
{\rm Im} F(s+i0)|_{s \in (s_{\rm max},\infty)}$. To implement this idea,
we rewrite the dispersion relation (\ref{disprel}) in the following way:
\begin{equation}
F(s)
-\frac{f_{\pi}^2}{s}
-\frac{1}{\pi }\int_{s_{\rm max}}^{\infty }\frac{f(x)}{x-s}dx 
=\frac{1}{\pi }\int_{0}^{s_{\rm max} }\frac{f(x)}{x-s}dx \, . 
\label{disprel1}
\end{equation}
Then we can calculate the left-hand side of this equation in QCD, 
\begin{equation}
\widetilde{F}_{\rm QCD}(s) :=
F_{\rm QCD}(s)
-\frac{f_{\pi}^2}{s}
-\frac{1}{\pi }\int_{s_{\rm max}}^{\infty }\frac{f_{\rm QCD}(x)}{x-s}dx 
\, ,
\end{equation}
whereas the right-hand side can be determined from experimental data.
Thus we shall test the hypothesis whether the left-hand side of
(\ref{disprel1}) can be described by QCD while the right-hand side
describes the experimental data.

In order to do that, we need an {\em a-priori} estimate of the accuracy
of the QCD predictions. This can be described by a continuous, strictly
positive function $\sigma_{L}(s)$ for $s \in \Gamma_{L}$ which should
encode errors due to the truncation of the perturbative series and the
operator product expansion. It is expected to decrease as
$|s|\rightarrow \infty $ and diverge for $s\rightarrow 0$. In Ref.\ 
\cite{CSSS} we aimed at a determination of the $D=6$ condensate of
$\Pi_{\rm V-A}$. There we were allowed to consider the contribution of
the dimension $D=8$ operator as an error, using $\sigma_{L}(s) = |{\cal
  O}^{\rm V-A}_{8}|_{\rm max}/s^{4}$ with $|{\cal O}^{\rm V-A}_{8}|_{\rm
  max}$ in the order of $10^{-3} \, {\rm GeV}^{8}$. If the perturbative
part dominates, as is the case for the individual vector or axial vector
correlators, the last known term of the perturbation series could be
used as a sensible estimate of the error corridor, possibly combined
with the first omitted term in the series over condensates. In the
present work, we focus on obtaining information on the correlation of
the condensates ${\cal O}^{\rm V-A}_6$ and ${\cal O}^{\rm V-A}_8$;
therefore we will use in a similar way an estimate of $|{\cal O}^{\rm
  V-A}_{10}|_{\rm max}$ to define an error corridor which, in this case,
scales with $1/s^5$.

The goal is to check whether there exists any function $F(s)$ with the
above analyticity properties, the true amplitude, which is in accord
with both the data in $\Gamma _{{\rm exp}}$ and the QCD model in
$\Gamma_{L}$. In order to quantify the agreement we will define
functionals $\chi _{L}^{2}[f]$ and $\chi _{R}^{2}[f]$ using an $L^{2}$
norm. For the time-like interval we simply compare the true amplitude
$f(s)$ with the data and use the covariance matrix of the experimental
data as a weight function:
\begin{equation}
\chi_{R}^{2}[f] = 
\frac{1}{|\Gamma_{\rm exp}|}
\int_{s_{0}}^{s_{\rm max}} dx 
\int_{s_{0}}^{s_{\rm max}} dx^{\prime}\,
V^{-1}(x,x^{\prime }) \left(f(x)-f_{{\rm exp}}(x)\right) 
  \left( f(x^{\prime})-f_{{\rm exp}}(x^{\prime })\right) \,.
\label{chir2}
\end{equation}
Experimental data correspond to cross sections measured in bins of $s$,
so that we can calculate this integral in terms of a sum over data
points. The ALEPH data which we use are given for 140 equal-sized bins
of width $\Delta s = 0.025$ GeV$^2$ between $0$ and $3.5$ GeV$^2$.
$\chi_R^2$ given in (\ref{chir2}) is in fact the conventional definition
of a $\chi^{2}$ normalized to the number of degrees of freedom and has a
probabilistic interpretation: for uncorrelated data obeying a Gaussian
distribution we would expect to obtain $\chi_{R}^{2}=1$.  Since
experimental data at different energies are correlated, we instead
expect
\begin{equation}
\chi_{{\rm exp}}^{2} = 
\frac{1}{N} \sum_{i,j} \sqrt{V(s_{i},s_{i}) V(s_{j},s_{j})}
V^{-1}(s_{i},s_{j}) \, .
\end{equation}

In order to define a measure for the agreement of the true function
$f(s)$ with theory, we use the left-hand side of (\ref{disprel1}) which
is well-defined and expected to be a reliable prediction of QCD in the
space-like interval for not too small $|s|$. This expression can be
compared with the corresponding integral over the true function. Thus we
define
\begin{equation}
\chi^2_L[f] = 
\frac{1}{|\Gamma_L|}
\int_{\Gamma_L} w_L(x) 
\left( \widetilde{F}_{{\rm QCD}}(x) 
- \frac{1}{\pi} \int_{s_0}^{s_{\rm max}} 
\frac{f(x^{\prime})}{x^{\prime}-x}dx^{\prime}\right)^2 dx
\label{chil2}
\end{equation}
where $w_L$ is the weight function for the space-like interval and
identified with $1/\sigma_L^2(s)$.

In order to find a best estimate of the function $f(s)$, we minimize
$\chi_L^2$ subject to the condition $\chi^2_R[f] \le \chi^2_{\rm exp}$.
The solution can be obtained by solving a Fredholm equation of the
second kind. This can be done numerically by expanding $f(s)$ in terms
of Legendre polynomials (see Ref.\ \cite{CSSS} for details).  The
algorithm to determine an acceptable value for the condensate is then
the following:
\begin{itemize}
\item[\bit i)] For the given value of $\chi^2_{\rm exp}$ we determine
  the solution for $f(s)$ iteratively and calculate the corresponding
  value of $\chi_L^2[f]$ as a function of ${\cal O}_D^{\rm V-A}$.
\item[\bit ii)] We minimize this $\chi_L^2[f]$ with respect to the
  values of ${\cal O}_D^{\rm V-A}$. 
\item[\bit iii)] We determine the error on the condensates by solving
  $\chi _{L}^{2} = \chi _{L,{\rm min}}^{2} + \Delta \chi^2$, where
  $\Delta \chi^2$ is chosen in the conventional way to reflect 1 or
  2$\sigma$ confidence regions for the values of ${\cal O}_D^{\rm V-A}$.
\end{itemize}

The size of $\sigma_L$ determines the minimal value of $\chi_L^2$
according to Eq.\ (\ref{chil2}) that can be reached by this algorithm.
Obviously, widening the error corridor (increasing $\sigma_L$) will lead
to values for $\chi_{L,{\rm min}}^2$ as small as desired. In such a
case, the information obtained from the fit is not conclusive, since any
model function $f(s)$ can be made consistent with the data if one allows
for a wide enough error corridor. On the other hand, narrowing the error
corridor will increase $\chi_{L,{\rm min}}^2$, signalling a bad fit,
i.e.\ bad consistency of theory with data if the model function is not
perfectly describing the data. However, a non-trivial result of our
approach is the fact, that there exists a choice for the error corridor
that leads to values $\chi_{L,{\rm min}}^2 = O(1)$. We shall assume that
the underlying probability distribution is Gaussian, and choose
$\chi_{L,{\rm min}}^2$ and accordingly $\sigma_L$ such that the fit
result corresponds to a $1\sigma$ CL. In practice this is done by
adjusting $|{\cal O}^{\rm V-A}_8|_{\rm max}$ (for the 1-parameter fit of
${\cal O}^{\rm V-A}_6$) or $|{\cal O}^{\rm V-A}_{10}|_{\rm max}$ (for
the 2-parameter fit of ${\cal O}^{\rm V-A}_6$ and ${\cal O}^{\rm
  V-A}_8$).
 

\section{Numerical results and discussion}

\begin{figure}[t!]
\centering
\includegraphics[height=.44\textwidth,angle=0]{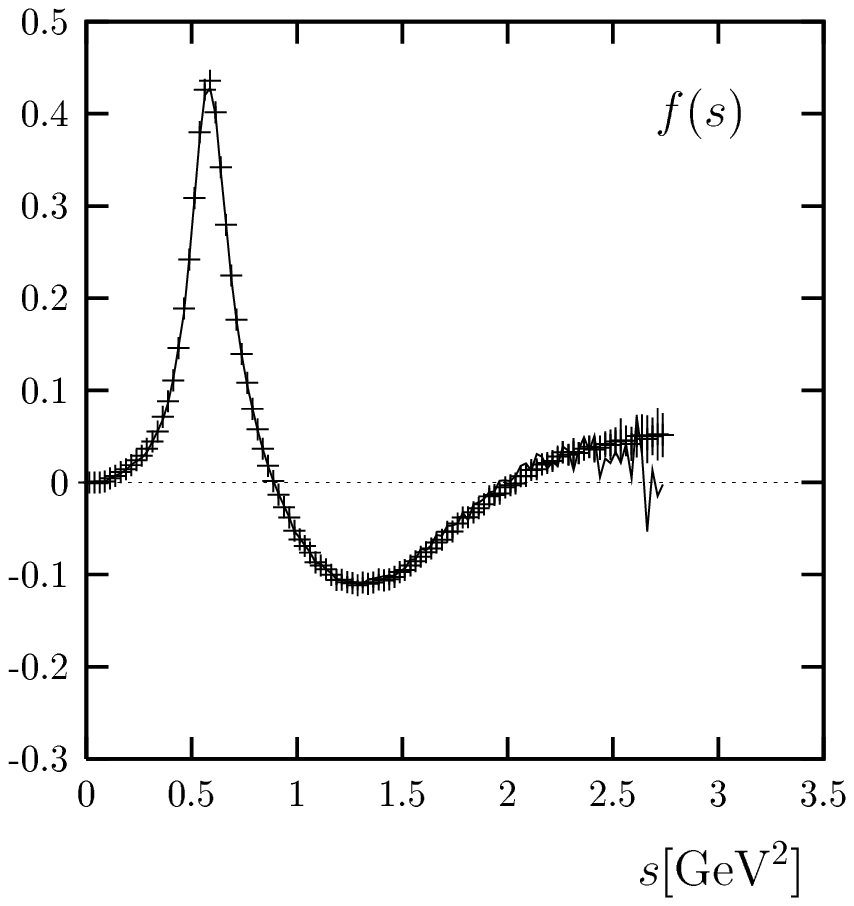}
\includegraphics[height=.44\textwidth,angle=0]{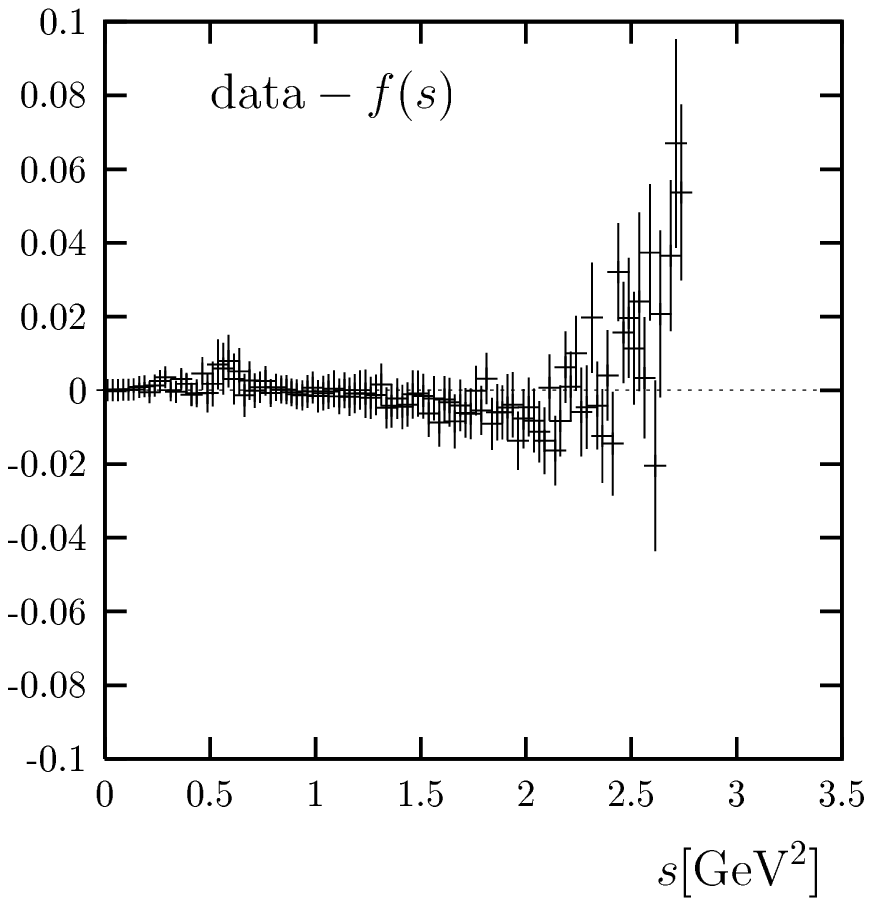}
\caption{The regularised function $f(s)$ compared with ALEPH data
  \protect\cite{aleph-final} (left). The right figure shows the
  difference.} 
\label{fig1}
\end{figure}

We start with a discussion of results obtained by 1-parameter fits of
the dimension $D=6$ condensate. We find a consistent description of the
data by QCD predictions including a non-zero value of ${\cal O}^{\rm
  V-A}_6$ as in our previous publication. The quality of the fit is
good, i.e.\ a value of $\chi_{L,{\rm min}}^2$ corresponding to a
$1\sigma$ CL can be fixed by choosing an error corridor described by the
contribution from an ${\cal O}^{\rm V-A}_8$ condensate of the expected
size of $\simeq 1.3 \cdot 10^{-3}$ GeV$^8$.  A direct comparison of the
data with the regularised function $f(s)$ (see Fig.\ \ref{fig1}) shows a
nice agreement over the full range of $s$ with the exception of the
highest $s$-bins, where experimental errors are large.  In comparison
with \cite{CSSS}, the agreement between data and theory has improved. A
number of consistency checks, as described in \cite{CSSS}, has been
performed and did not reveal any problem.

The results of the 1-parameter fit can be summarized by quoting values
for ${\cal O}_6^{\rm V-A}$ at LO and at NLO for the two available values
of the NLO coefficients\footnote{Note that the normalization differs by
  a factor of 2 with respect to \cite{CSSS}}:
\begin{eqnarray}
{\cal O}_6^{\rm V-A} & = & -5.9^{+1.7}_{-1.0} \times 10^{-3}
{\rm GeV}^6 
~~~{\rm for}~~~\tilde{c}_6 = 0, ~~ \mbox{LO} \, ,
\label{oneparfLO} \\
{\cal O}_6^{\rm V-A} & = & -4.9^{+1.5}_{-1.1} \times 10^{-3}
{\rm GeV}^6 
~~~{\rm for}~~~\tilde{c}_6 = \frac{89}{12}\, ,
\label{oneparf89} \\
{\cal O}_6^{\rm V-A} & = & -3.6^{+1.2}_{-1.2} \times 10^{-3}
{\rm GeV}^6 
~~~{\rm for}~~~\tilde{c}_6 = \frac{247}{12}\, .
\label{oneparf247}
\end{eqnarray}
The NLO results are based on the 4-loop expression for $\alpha_s$ with
$\Lambda_{\overline{{\rm MS}}}(N_f=3) = 0.326$ GeV and the
renormalization scale chosen equal to $|s|$. They are not sensitive to
changing $\Lambda_{\overline{{\rm MS}}}$ within the present experimental
error of $\pm 0.030$ GeV. Moreover, the fit results based on the two
different values of the NLO coefficients $\tilde{c}_6$ agree within
errors and their difference with respect to the LO result is consistent
with a shift calculated from the correction term choosing a typical
value of $O(1.5)$ GeV for the renormalization scale in $\alpha_s$. This
is to be expected since our method would work for any $s$-dependent
ansatz for ${\cal O}^{\rm V-A}_6$ as well.

\begin{figure}[t!]
\centering
\includegraphics[height=.6\textwidth,angle=0]{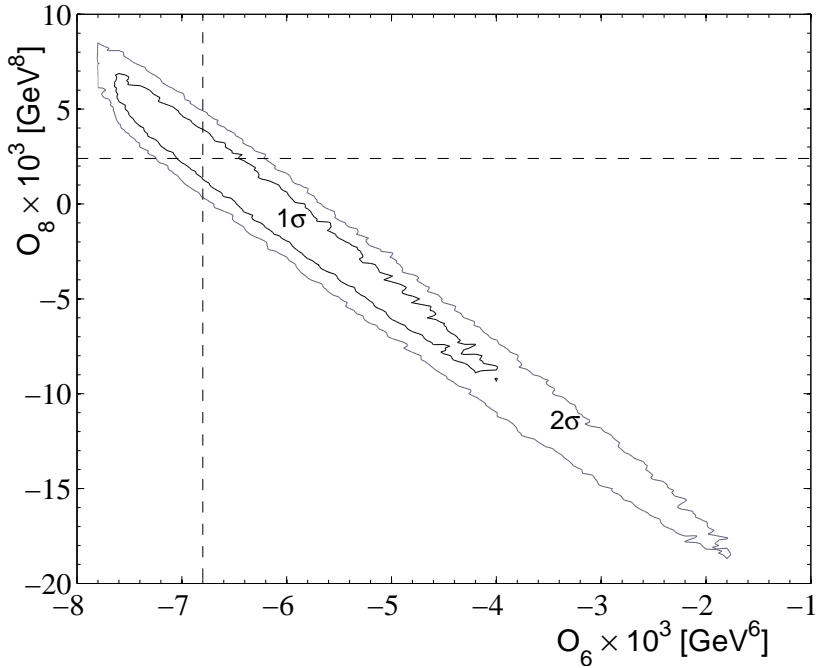}
\caption{1- and $2\sigma$ CL regions in the $({\cal O}_6^{\rm V-A},
  {\cal O}_8^{\rm V-A})$-plane from the 2-parameter fit (LO). The
  central values are marked by the dashed lines.}
\label{fig2}
\end{figure}

The main result of our present analysis is obtained from a 2-parameter
fit of the dimension $D=6$ and $D=8$ condensates. Here we do not include
the NLO contribution to ${\cal O}_6^{\rm V-A}$ since the corresponding
NLO coefficient for ${\cal O}_8^{\rm V-A}$ is not known. In this case we
find agreement of theory and data at the $1\sigma$ CL if we choose an
error corridor described by the dimension $D=10$ contribution with the
value $\simeq 5.7 \cdot 10^{-3}$ GeV$^{10}$. The result presented in
Fig.\ \ref{fig2} shows a strong negative correlation of ${\cal O}^{\rm
  V-A}_6$ and ${\cal O}^{\rm V-A}_8$. Both the central values as well as
the errors from the 2-parameter fit are consistent with those from the
LO 1-parameter analysis: The $1\sigma$ range allowed for ${\cal O}^{\rm
  V-A}_6$ for fixed ${\cal O}^{\rm V-A}_8 = 0$ (which is the assumption
underlying the 1-parameter fit) agrees with (\ref{oneparfLO}).  However,
leaving the value of ${\cal O}^{\rm V-A}_8$ unconstrained, as in the
2-parameter fit, one finds a much larger range for ${\cal O}^{\rm
  V-A}_6$.  The minimum value of $\chi_L^2$ for the 2-parameter fit is
located at the values
\begin{eqnarray}
{\cal O}_6^{\rm V-A} & = & 
-6.8_{-0.8}^{+2.8} \times 10^{-3}\mbox{GeV}^6\, ,
\nonumber \\
{\cal O}_8^{\rm V-A} & = & 
2.4_{-11.7}^{+4.5} \times 10^{-3}\mbox{GeV}^8 \, .
\label{twoparf}
\end{eqnarray}
The errors on ${\cal O}_6^{\rm V-A}$ for fixed ${\cal O}_8^{\rm V-A}$
are small, but the allowed range for ${\cal O}_8^{\rm V-A}$ is not very
restrictive (note the different scales for the two condensates in Fig.\ 
\ref{fig2}). However, the strong correlation allows one to determine a
linear combination of ${\cal O}_6^{\rm V-A}$ and ${\cal O}_8^{\rm V-A}$
with a rather small error:
\begin{equation} 
{\cal O}_8^{\rm V-A}+3.04~\mbox{GeV}^2 \cdot{\cal O}_6^{\rm V-A}
=
-18.30^{-0.25}_{+0.38}\times 10^{-3} \mbox{GeV}^8\, .
\end{equation}
We also checked that we obtain consistent results when including the NLO
correction to ${\cal O}_6^{\rm V-A}$: The 1- and $2\sigma$ contours are
shifted, essentially without changing their form, to larger values of
${\cal O}_6^{\rm V-A}$ exactly as can be inferred from the 1-parameter
fits: for $\tilde{c}_6 = 0$ ($89/12, 247/12$) we find the minimum of
$\chi_L^2$ at ${\cal O}_6^{\rm V-A} \times 10^3 / \mbox{GeV}^6 = -6.8~
(-6.6, -5.8)$.

There exist a number of previous extractions of QCD condensates in the
literature, all based on a sum rule approach and at LO
\cite{ZyabD10,DGHS,BGP,IZ,Golo,CAD+KS,RoLa,Narison,BDPS}. The results
for ${\cal O}_6^{\rm V-A}$ cover values between $(-2.27 \pm 0.51) \cdot
10^3$ GeV$^6$ \cite{Golo} and $(-8.0 \pm 2.0) \cdot 10^3$ GeV$^3$
\cite{CAD+KS}. These values correspond to a scale of about 200 MeV which
is comparable to $\Lambda_{\rm QCD}$. Although errors given by the
authors are typically in the order of $25\,\%$, their central values are
only in rough agreement.  The observed variations of these results
represent the ambiguities inherent in the QCD sum rule approach. Our
result nicely falls into the same range, also with an error estimate of
the same size.

For ${\cal O}_8^{\rm V-A}$, previous results range from ($-10.8 \pm
6.6)\cdot 10^{-3}$ GeV$^8$ \cite{BDPS} to $(12.2 \pm 2.9)\cdot 10^{-3}$
GeV$^8$ \cite{Narison}. A recent conservative estimate \cite{RoLa} is
${\cal O}_8^{\rm V-A} = (-12^{+7}_{-11})\cdot 10^{-3}$ GeV$^8$. Again,
our result agrees within the estimated precision.  Errors for the $D=10$
condensate, ${\cal O}_{10}^{\rm V-A}$, are typically larger and the
spread of values found in the literature even larger: they range from
$(-18.2 \pm 5.9) \cdot 10^{-3}$ GeV$^{10}$ \cite{Narison} to $(78 \pm
24)\cdot 10^{-3}$ GeV$^{10}$ \cite{RoLa}. These values are consistent
with the value we had to choose for the error corridor to determine the
$1\sigma$ CL range in the 2-parameter analysis. Moreover, it is also
interesting to note the agreement of the correlation between ${\cal
  O}_6^{\rm V-A}$ and ${\cal O}_8^{\rm V-A}$ with corresponding results
from \cite{ZyabD10,Narison}.

Obviously, our numerical results are, from a practical point of view,
not superior to approaches based on finite energy sum rules. However,
the fact that we find agreement within errors is not trivial. Since our
approach is based on much more general assumptions, the results obtained
in this analysis give additional confidence to the numerical values
obtained with the help of QCD sum rules.


\section*{Acknowledgment}

We are grateful to A.\ H\"ocker for providing us with the ALEPH data and
for helpful discussions.




\end{document}